# Parameter Calibration in Crowd Simulation Models using Approximate Bayesian Computation

**Nikolai W.F. Bode[1]**

[1]Department of Engineering Mathematics, University of Bristol
Merchant Venturers Building, BS8 1YB, Bristol, UK
nikolai.bode@bristol.ac.uk

*Abstract* - Simulation models for pedestrian crowds are a ubiquitous tool in research and industry. It is crucial that the parameters of these models are calibrated carefully and ultimately it will be of interest to compare competing models to decide which model is best suited for a particular purpose. In this contribution, I demonstrate how Approximate Bayesian Computation (ABC), which is already a popular tool in other areas of science, can be used for model fitting and model selection in a pedestrian dynamics context. I fit two different models for pedestrian dynamics to data on a crowd passing in one direction through a bottleneck. One model describes movement in continuous-space, the other model is a cellular automaton and thus describes movement in discrete-space. In addition, I compare models to data using two metrics. The first is based on egress times and the second on the velocity of pedestrians in front of the bottleneck. My results show that while model fitting is successful, a substantial degree of uncertainty about the value of some model parameters remains after model fitting. Importantly, the choice of metric in model fitting can influence parameter estimates. Model selection is inconclusive for the egress time metric but supports the continuous-space model for the velocity-based metric. These findings show that ABC is a flexible approach and highlight the difficulties associated with model fitting and model selection for pedestrian dynamics. ABC requires many simulation runs and choosing appropriate metrics for comparing data to simulations requires careful attention. Despite this, I suggest ABC is a promising tool, because it is versatile and easily implemented for the growing number of openly available crowd simulators and data sets.

## 1. Introduction

Simulation models for pedestrian crowds are a widely used tool [1]. The dynamics these models produce are controlled by parameters that capture the preferred speed of pedestrians or the strength of interactions between pedestrians, for example [1]. As many models are intended to be used or are already used to investigate real world scenarios, a key challenge is to calibrate model parameters, such that simulations produce realistic behaviour [2]. In addition, robust approaches for calibrating parameters facilitate a fair comparison of the predictive potential or goodness of fit across different models [3].

A range of approaches for calibrating model parameters have been suggested [2,4-8]. They compare models to empirical data at a microscopic level (e.g. trajectories [5]) or at a macroscopic level, where summary statistics for simulations and data are compared (e.g. pedestrian flows [2]). For specific models, it is sometimes possible to formulate a likelihood function linking model and data via probability distributions [6]. However, for most simulation models it is not practical to find explicit, closed-form probability distributions for simulation outcomes. Thus, parameter calibration typically uses an objective function that measures the difference between data and simulations via measures derived from microscopic data [5,7,8]. Parameter estimates are found by optimising the objective function. While this approach is valid, it has three major shortcomings. First, this approach yields point-estimates for parameters and provides no information on the uncertainty associated with estimates. Second, this approach is not suited for model comparison, where the relative quality of different models in describing data is established. The objective function does provide a goodness of fit measure, but it is not clear how differences in model complexity (i.e. number of model parameters) should be accounted for when comparing this measure across models. Third, all numerical optimization procedures are liable to getting stuck in local optima, meaning that the true optimal solution may not be found.

In this contribution, I propose an alternative approach for pedestrian model calibration using Approximate Bayesian Computation (ABC), that is already widely used in other fields of science (e.g. [9]). I show how this flexible framework avoids the issues mentioned above and I demonstrate for representative simulation models that even for simple scenarios (e.g. unidirectional flow through one bottleneck), parameter estimates are associated with substantial uncertainty.

## 2. Methods

To demonstrate the parameter calibration and model selection approach based on Approximate Bayesian Computation (ABC), I consider two different simulation models and two different approaches for comparing simulations to data. In the following, I first describe the two models, then I outline the ABC approach and finally I provide details on model fitting and selection.

### 2.1. Microscopic Pedestrian Simulation Models

I consider two microscopic models that are derived from popular models in the literature. As this work is intended as a demonstration of principle only, I do not wish to make any claims about the overall quality of either model or similar models.

The first model is a derivative of a popular simulation model that describes the movement of pedestrians in continuous space [10]. To save space, I do not provide a detailed description of this model, but all details can be found in precious work [11]. Briefly, pedestrian-pedestrian and pedestrian-wall interactions are captured in force vectors acting on simulated pedestrians. The movement preferences of simulated pedestrians, e.g. towards a target and away from walls, is encoded in a discrete floor field [11]. Models of this type are commonly referred to as "Social Force models" and I thus refer to this model as "SF model" here. I fit the values of five parameters of this model, keeping the remaining parameters fixed at default levels that have been specified previously [11]. The fitted parameters are the preferred speed of pedestrians, $v^0$, coefficients determining the strength of psychological interactions between pedestrians $(A, B)$ and coefficients regulating the strength of physical pedestrian-pedestrian and pedestrian-wall interactions $(k, \kappa)$. The nomenclature for these parameters is identical to the one used in previous work [11].

The second model is a cellular automaton model that describes the movement of pedestrians in discrete space on a lattice grid with square cells of side-length 0.4m. It is very loosely based on previous work [12]. The model is deliberately kept very simple to provide a contrast to the complex SF model. Let $F_{ij}$ denote the value of cells of a static floor field that encodes the movement preferences of pedestrians in the same way as the floor field used for the SF model. The same algorithm as for the SF model is used to construct $F_{ij}$ (see [11]). The indices $i$ and $j$ denote the rows and columns of the lattice grid. Furthermore, let $M_{ij}$ be a matrix that takes value 1 if cell $(i, j)$ of the lattice grid that pedestrians move on is occupied by a pedestrian or a wall and 0 otherwise. The positions of all simulated pedestrians are updated synchronously in fixed time steps of length $\delta t$ seconds. Simulated pedestrians move to cells within the Moore neighbourhood of their current position according to probabilities $P_{ij}$ that are computed as follows:

$$P_{ij} = \frac{([1 + F_{ij}][1 - M_{ij}])^S}{C} \qquad (1)$$

$S$ is a model parameter that captures how strongly simulated pedestrians respond to gradients in the floor field $F_{ij}$ and $C$ is a normalising constant that ensures the $P_{ij}$ sum to 1 over the Moore neighbourhood for any given simulated pedestrian. Eq. (1) ensures that simulated pedestrians cannot move onto walls or grid cells that are already occupied by other pedestrians. If the procedure above leads to a conflict between two or more simulated pedestrians moving onto the same grid cell, only the simulated pedestrian with the highest value of $P_{ij}$ associated with this cell is moved and the others remain stationary. I fit two parameters of this model: $S$ and $\delta t$. The former parameter controls how strongly simulated pedestrians respond to gradients in the floor field and the latter parameter controls the update rate and thus the maximum attainable speed of simulated pedestrians. As this model is a cellular automaton, I refer to it as "CA model".

The simulation scenario I consider throughout is one where 70 simulated pedestrians exit a square 10mx10m room through a single bottleneck that is 1.2m, 1.6m or 2.0m wide and 0.4m deep. Pedestrians are considered to have left the room once they have passed through the bottleneck and simulated pedestrians are removed from the simulation once they have moved 1.0m away from the bottleneck after leaving the room.

The next section describes the general principles of the model fitting approach before section 2.3 details how the SF and the CA model are fitted to data.

## 2.2. Approximate Bayesian Computation

ABC is a likelihood-free statistical model fitting technique [9]. Instead of an explicit likelihood function, model simulations are used in the model fitting procedure. In its simplest form, ABC consists of five steps that are described below (Fig. 1).

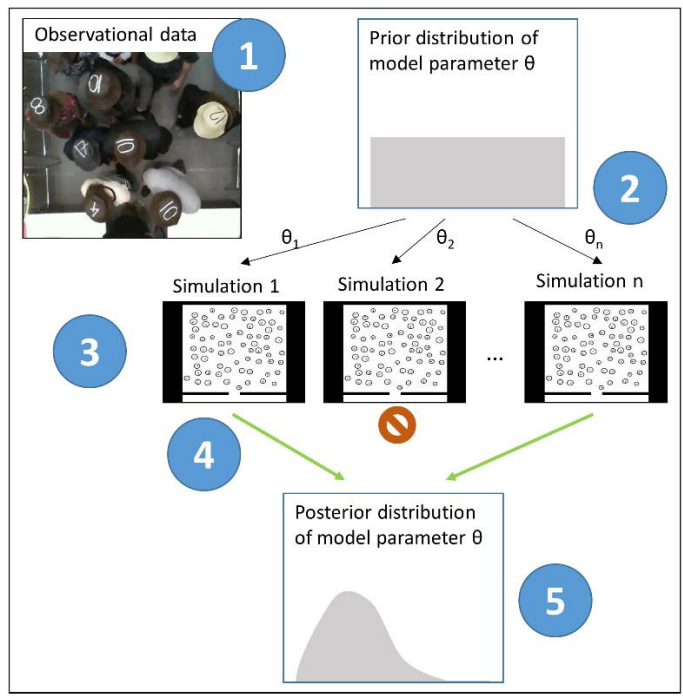

Fig. 1 Illustration of Approximate Bayesian Computation. Steps 1-5 labelled in the figure are explained in the text.

(1) Obtain measures of interest from data. These could be microscopic or macroscopic measures.
(2) For a given model, decide on a prior distribution for model parameters, $\theta$. In Fig. 1 we assume no prior knowledge about $\theta$ and therefore use uniform or flat distributions. $\theta$ can be a vector including multiple parameters.
(3) Draw $n$ samples, $\theta_1$ to $\theta_n$ from the prior distribution and use them to perform model simulations.
(4) Compute the measure of interest from (1) for simulations and use a distance function and a tolerance $\varepsilon$ to decide if a model simulation is sufficiently similar to the data.
(5) Approximate the posterior distribution of θ with the parameter values accepted in (4) – $\theta_1$ and $\theta_n$ in Fig. 1.

This processs works for different measures of interest (e.g. pedestrian flow, fundamental diagram, trajectories), as long as a suitable distance function linking data to simulation can be found (step 4 above). Provided the tolerance $\varepsilon$ is chosen small enough and enough simulations are performed (large $n$), ABC will find an accurate estimate of the posterior parameter distribution [9]. The posterior parameter distribution indicates both the most likely parameter values (e.g. mean or mode of posterior) and the uncertainty associated with it (e.g. variance of posterior). A particularly useful feature of ABC is that it can be used to compare the quality of models in explaining data. If ABC is performed for different models on the same data using the same value of $\varepsilon$, then the rate at which parameters are accepted into the posterior distribution (step 4) for each model can be used to approximate the Bayes Factor – a commonly used measure for model selection [9]. Model complexity, i.e. the number of parameters a model has, is inherently accounted for in ABC [9].

### 2.3. Model Fitting and Model Selection

For model calibration I focus on the simple benchmark scenario of unidirectional flow through one bottleneck, as already indicated above. For this scenario, experimental data including the trajectories of all pedestrians within a measurement area in front of the bottleneck is publicly available (data published in [13], available on http://ped.fz-juelich.de/db/). As already described above, I consider bottleneck widths of 1.2m, 1.6m and 2.0m.

The model fitting procedure requires a measure recorded from the experimental data that can be used to compare data and simulations. I use two different measures to investigate the robustness of the model fitting procedure.

The first measure I use is the time it takes 30 pedestrians to pass through the bottleneck after the first 10 pedestrians have exited, denoted $\Delta T_{10}^{40}$ (Fig. 2a). The measured times for the experimental data are 9.68s, 7.64s and 5.20s for the bottleneck widths 1.2m, 1.6m and 2.0m, respectively. This measure presents a crude approximation of the pedestrian flow. I focus on data close to the start of experiments and simulations to avoid artefacts caused by transient effects, such as changes in behaviour at the end of experiments. For comparison with simulations, I compute the Euclidean distance between the times obtained from the data and from simulations.

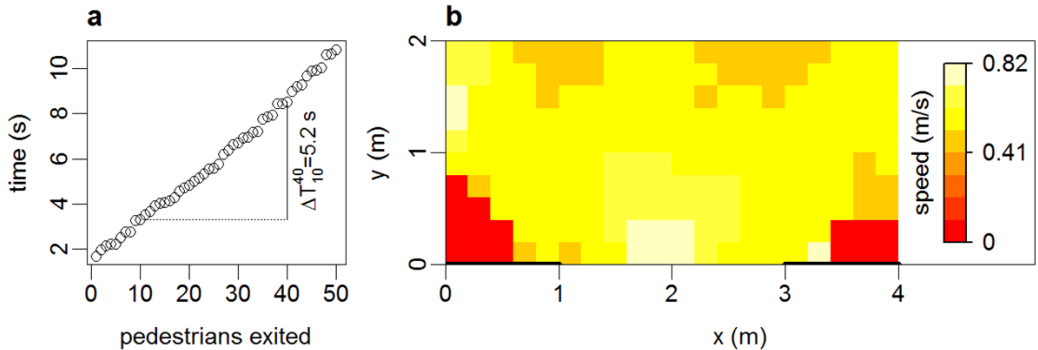

Fig. 2: Measures used to compare experimental data to simulations. Panel (a) shows the time it takes 30 pedestrians to pass through the bottleneck, after the first 10 pedestrians have exited. Panel (b) shows the field of the average speed directed at the bottleneck centre in a measurement area in front of the bottleneck. The data shown is for a bottleneck width of 2m. The centre of the bottleneck is located at $(x, y) = (2, 0)$. See text for details.

The second measure I use is recorded over the same time interval as the first measure (i.e. between the 10[th] and the 40[th] pedestrian passing through the bottleneck). This measure is inspired by work which suggests that the variation of average pedestrian velocities at different locations in front of bottlenecks is informative for the overall dynamics [14]. It is measured on a 4m by 2m grid located in front of the bottleneck that consists of square cells of side-length 0.2m. For each grid cell, individuals located inside the cell are considered and the average of the component of their velocity that is directed towards the centre of the bottleneck is recorded (Fig. 2b). In other words, the measure captures how fast pedestrians are moving towards the centre of the bottleneck at different locations in front of the bottleneck. Fig. 2b shows a V-shaped funnel of this speed field, suggesting that movement speeds towards the bottleneck are not homogeneous in space. For comparison with simulations, I compute the Euclidean distance between speed fields obtained from experimental data and from corresponding simulations and average over the three bottleneck widths.

For both distance measures, I fit the two models by performing simulations with 70 pedestrians for 1,000,000 samples from the prior distribution for each bottleneck width. For the distance measure based on exit times I then use a tolerance of $\varepsilon = 2s$ to obtain posterior parameter distributions. For the distance measure based on the field of speeds directed towards the exit, I use a tolerance of $\varepsilon = 10 m/s$ to obtain posterior parameter distributions. I use flat priors (i.e. uniform distributions) for all parameters and the range of values for the prior distributions is detailed in section 3 below.

## 3. Results and Discussion

All posterior parameter distributions for the SF model differ substantially from the uniform prior distributions (Fig. 3). This change between the prior and posterior parameter distributions indicates the information we gain about the parameter values from the data using ABC model fitting procedure. For all parameters and in particular for $A$ (Fig. 3a), the posterior distribution indicates that a substantial level of uncertainty associated with the estimate of the parameter remains (consider the spread of the distribution). Decreasing the ABC acceptance threshold $\varepsilon$ and increasing the number of simulations performed in ABC, may address these issues to some extent, but some uncertainty for parameter estimates will always remain, reflecting inherent variability in simulations.

Comparing the posterior distributions obtained from fitting the model using the two different distance measures between data and simulations (egress time and speed field) reveals that the choice of distance measure can have a substantial effect on the parameter estimates and therefore on model calibration. For example, while the estimate for parameter $k$ is robust to the choice of distance measure in the ABC model fitting (Fig. 3d), the mean of parameter $v^0$ is shifted with model fitting using the speed field leading to a lower estimate of $v^0$, on average. This shows that fitting the SF model to one feature of data does not necessarily lead to the best fit for other aspects of data (see e.g. discussion in [15] for a different model). This is not a surprising insight for complex models, but I suggest the ABC model fitting approach is useful for investigating such differences in several ways. First, the posterior parameter distributions indicate how parameter estimates differ for model fits that use different distance measures. Second, when using ABC, several distance measures can be combined, as desired, depending on the context. For example, multiple different measures could be combined in a weighted sum or otherwise to provide a new distance measure that takes more aspects of the data into account. Third, ABC provides a rigorous approach for comparing the quality of different models, depending on what distance measure is used (see below).

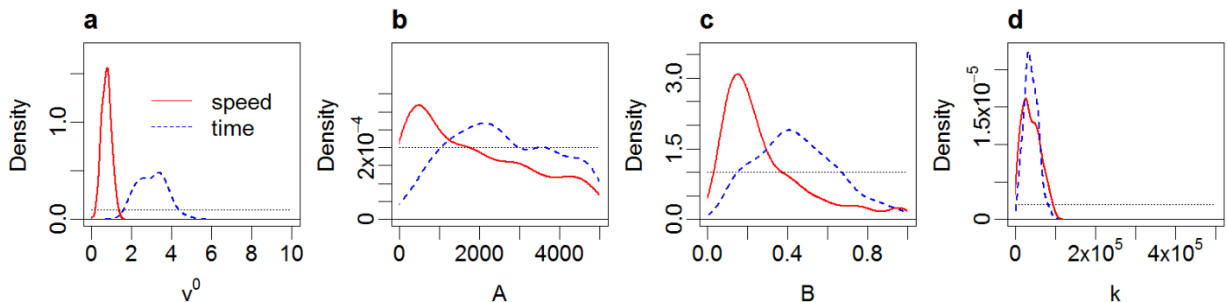

Fig. 3: Marginal posterior distributions for parameters in the SF model. The solid red line indicates the posterior obtained using the speed field distance measure, the dashed blue line shows the posterior obtained using the egress

time distance measure and the dotted horizontal line indicates the uniform prior distribution. Due to spatial constraints, I only report the marginal posterior distributions for four of the five fitted model parameters here.

ABC model fitting for the CA model provides qualitatively similar insights (Fig. 4). Posterior distributions indicate the evidence on parameters we gain from the data. The value of $\delta t$ is identified with high certainty (highly peaked posterior, Fig. 4a) for both distance measures, while we learn little about the value of $S$ when considering egress times (equivalently, $S$ does not affect egress times substantially), but there is comparatively little uncertainty about $S$ when considering the speed field (Fig. 4b). This also shows that as for the SF model, using different distance measures when fitting the CA model leads to different posterior distributions for parameters.

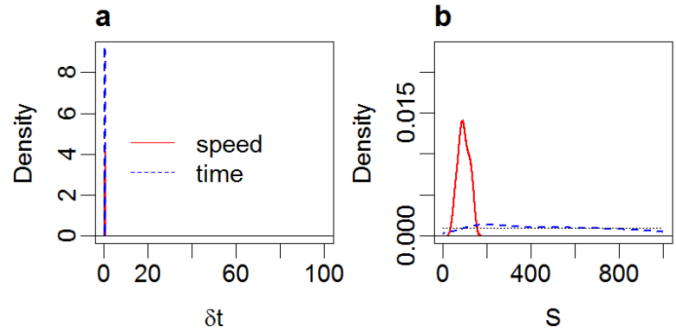

Fig. 4: Marginal posterior distributions for parameters in the CA model. The solid red line indicates the posterior obtained using the speed field distance measure, the dashed blue line shows the posterior obtained using the egress time distance measure and the dotted horizontal line indicates the uniform prior distribution.

When choosing acceptance thresholds $\varepsilon$ in ABC, values that are as small as possible while still leading to meaningful approximations of the posterior parameter distributions are desired. Table 1 shows that the acceptance rates for samples from the prior are low for both models, but Figs 3 and 4 show that posterior distributions can still be estimated. To compare the relative quality of the SF and CA model in explaining the data, the Bayes Factor between the two models can be approximated from the rate at which samples from the prior distribution are accepted into the posterior distribution in ABC [9]. This approach for comparing the quality of models implicitly takes model complexity (i.e. number of model parameters) into account. I compute the Bayes Factor (BF) between the SF and the CA model. Therefore, positive values of $log(BF)$ indicate support for the SF model and vice-versa. If the BF is close to zero, it is difficult to distinguish between the two models. I adopt a commonly used scale for interpreting $2log(BF)$ values which suggests that only absolute values larger than 2 indicate positive evidence for either model [16].

Table 1: ABC acceptance rates for the two models and the two distance measures between data and simulations used. The Bayes Factor (BF) can be approximated as the fraction of the two ABC acceptance rates [9]. I compute the BF between the SF and the CA model. Thus, positive values of log(BF) indicate support for the SF model.

|  | SF model | CA model | 2log(Bayes Factor) |
|---|---|---|---|
| **Acceptance rate egress time** | $2.21 \times 10^{-4}$ | $5.46 \times 10^{-4}$ | $-1.80$ |
| **Acceptance rate speed field** | $2.75 \times 10^{-4}$ | $2.20 \times 10^{-5}$ | $5.04$ |

Table 1 shows that fitting models using the egress time measure leads to a negative value of $2log(BF)$ which would indicate that the CA model is better supported by the data than the SF model. However, as the absolute value of $2log(BF)$ is lower than 2, the model selection using the egress time measure is inconclusive. In contrast, model fitting using the speed field measure leads to a to a positive value of $2log(BF)$ that is larger than 2, which indicates positive evidence in favour of the SF model.

These findings suggest that the CA model can capture egress times as well as the SF model, but that the SF model outperforms it in capturing the fine-scale patterns encoded in the speed field. More generally, the contrasting outcomes of model selection for the two measures I used in model fitting indicate that model selection does not always provide a clear outcome. It is possible that different models are suitable for describing different aspects of pedestrian dynamics. If this was the case, which model to select or prefer would then depend on the context and the intended use of the model. As already discussed above, it is also possible to combine different distance measures in ABC to take multiple features of data into account.

## 4. Conclusion

In summary, my contribution aims to demonstrate that ABC is a useful and flexible approach for pedestrian model calibration. ABC provides parameter estimates, indicates the uncertainty associated with the estimates and can be used for model selection. In addition, ABC is an excellent tool to highlight the difficulties of model calibration and model selection in pedestrian dynamics.

On the one hand, my findings demonstrate that ABC provides a flexible approach to fit models for pedestrian dynamics to data and to compare the relative quality of models for explaining data. On the other hand, it is important to note that the success of ABC relies on two main components. First, ABC requires a suitable quantitative measure to compare data and simulations. I have demonstrated the use of two different measures here. The differences in model fitting and selection results depending on the measure used highlight the importance of carefully deciding how to compare data and simulations. Second, ABC requires many numerical simulations of models. How many simulations are needed depends on the prior information available. When starting without much prior information (as I demonstrate here with flat priors), millions of simulations are needed. The availability of information, e.g. on a narrow range of values parameters can take, or even information on the most likely parameter values, will reduce the required number of simulations for ABC model drastically. For confined scenarios, like the bottleneck scenario I consider here, it is feasible to run large numbers of simulations. However, if little prior information is available and models are to be fitted to data from highly complex scenarios (e.g. involving many pedestrians or entire buildings), the high number of simulations required may become prohibitive. One way to approach such scenarios could be to initially fit models to subsets of the scenario investigated (e.g. only considering part of a building or crowd) or ideally to separate data and to subsequently use the posterior parameter distributions from this model fitting as prior distributions in model fitting.

In addition to being flexible, I suggest ABC is a useful addition to the model calibration and model selection efforts in pedestrian dynamics because once a simulator for a given model is available, it is very easily implemented. As model simulators are increasingly being made openly available [17-19], I believe now is a good time to introduce ABC into the field of pedestrian modelling.